# Measuring foetal dose from tomotherapy treatments

Samuel C. Peet[1,2], Tanya Kairn[1,2], Craig M. Lancaster[1], Jamie V. Trapp[2], Steven R. Sylvander[1], Scott B. Crowe[1,2]

[1]Royal Brisbane and Women's Hospital, Butterfield Street, Herston, QLD 4029, Australia

[2]School of Chemistry, Physics and Mechanical Engineering, Queensland University of Technology, GPO Box 2434, Brisbane, QLD 4001, Australia

Corresponding author: Samuel Peet (samuel.peet.physics@gmail.com)

# Abstract

**Introduction** Treating pregnant women in the radiotherapy clinic is a rare occurrence. When it does occur, it is vital that the dose received by the developing embryo or foetus is understood as fully as possible. This study presents the first investigation of foetal doses delivered during helical tomotherapy treatments.

**Materials & Methods** Six treatment plans were delivered to an anthropomorphic phantom using a tomotherapy machine. These included treatments of the brain, unilateral and bilateral head-and-neck, chest wall, and upper lung. Measurements of foetal dose were made with an ionisation chamber positioned at various locations longitudinally within the phantom to simulate a variety of patient anatomies.

**Results** All measurements were below the established limit of 100 mGy for a high risk of damage during the first trimester. The largest dose encountered was 75 mGy (0.125% of prescription dose). The majority of treatments with measurement positions less than 30 cm fell into the range of uncertain risk (50 – 100 mGy). All treatments with measurement positions beyond 30 cm fell into the low risk category (< 50 mGy).

**Conclusions** For the cases in this study, tomotherapy resulted in foetal doses that are at least on par with, if not significantly lower than, similar 3D conformal or intensity-modulated treatments delivered with other devices. Recommendations were also provided for estimating foetal doses from tomotherapy plans.

# Introduction

Approximately one in 1,000 pregnancies are complicated by cancer [1]. In cases where patients are known to be pregnant and unlikely to survive the term of the pregnancy without radiotherapy treatment, the embryo or foetus may be deliberately exposed to a radiation dose [2]. Alternatively, in cases where patients assert that they are not pregnant at the time of radiotherapy treatment and then subsequently find that they were pregnant during the treatment, the embryo or foetus may be unintentionally exposed to a radiation dose. In both cases, it is necessary to provide the prescribing physician with an estimate of the radiation dose to the embryo or foetus and an evaluation of the risks involved in the treatment. The effects of radiation on an embryo or foetus depend on the stage of the pregnancy, and may involve organ malformation, growth retardation, reduction in mental capacity, induction of cancer, or death [3].

The American Association of Physicists in Medicine (AAPM) TG-36 report includes a set of guidelines for categorising the risk to the embryo or foetus as a function of dose [4]. The TG-36 report states that below 50 mGy, there is little risk of damage. Between 50 and 100 mGy, the risk is uncertain. From 100 to 500 mGy, there is significant risk of damage during the first trimester, and beyond 500 mGy there is a high risk during all trimesters. The report also provides a series of reference data for estimating foetal doses based on simple square fields from conventional linear accelerators. However, these data are not well suited to contemporary radiotherapy technologies such as helical tomotherapy, as noted in the recent AAPM report on out-of-field dose [5].

For any given treatment, helical tomotherapy units such as the Accuray TomoTherapy Hi-Art (Accuray, Sunnyvale, USA) deliver many times more monitor units than conventional linear accelerators. To combat the higher head leakage due to this increased machine output, the jaws are made thicker and the accelerator head is heavily shielded. This shielding is effective, even reducing the total peripheral dose to a lower value than conventional accelerators under certain conditions [6].

There are no published investigations of foetal doses in pregnant patients treated with tomotherapy. However, several studies of the peripheral dose delivered during clinical tomotherapy treatments do exist. One study delivered a single parotid-sparing head-and-neck treatment to a solid water phantom [6]. The peripheral dose was measured inferiorly from the field edge to a distance of 30 cm, where it was found that the dose dropped to 0.2% of the prescribed dose. Measurements were made at depths of 1.5 cm, 5 cm, and 10 cm, and the peripheral dose was found to be independent of depth. Another study performed a single measurement of peripheral head leakage during a nasopharynx treatment that was modified such that the multi-leaf collimator (MLC) leaves remained closed throughout the delivery [7]. The measurement was performed 20 cm from the isocentre towards the foot of the couch and was found to be 0.2% of the prescription dose. Another investigation characterised the internal and external scattering components of the peripheral dose by delivering a tomotherapy treatment to a simple cylindrical target volume in a perspex cuboidal phantom [8]. An alternative study [9] delivered a similar plan to a solid water slab phantom and performed a series of measurements out to 34 cm out-of-field, however, when several of these measurements were repeated with an anthropomorphic phantom the results were substantially different, indicating a dependence on patient geometry. A further study investigated tomotherapy irradiations of cardiac devices, measuring a dose at a specific location during a head-and-neck treatment (0.7% of the prescribed dose, 9 cm inferior to the planning target volume [PTV]), and during a chest wall treatment (27% of the prescribed dose, 9 cm lateral to the PTV) [10].

There are few published studies adequately describing peripheral doses in clinical treatment scenarios, nor are there any investigations specifically considering foetal doses.

Without this knowledge, an informed judgement of treating pregnant patients with tomotherapy is made difficult. Therefore, in this study we determined the magnitude of the peripheral dose delivered in the vicinity of an embryo or foetus in a range of clinical tomotherapy scenarios. The results may be used to aid in clinical decision making and when performing dose estimations.

## Method

### Treatment Plans

Six treatment plans were selected for delivery, encompassing a broad range of sites often treated with tomotherapy. A summary of the plans can be seen in Table 1. Brain 1 treated a large volume in the left temporal lobe (364 cm$^3$), included sparing of the brain stem, and was delivered with a 2.5 cm slice width. In contrast, Brain 2 treated a much smaller anterior clinoidal meningioma (12 cm$^3$) with a slice width of 1.0 cm. Two head-and-neck (H&N) cases were also selected: H&N 1 treated a squamous cell carcinoma of the left maxilla with the PTV also encompassing the nodes of the left neck; H&N 2 treated a base of tongue lesion and included bilateral nodes. Both plans were delivered with a 2.5 cm slice width, although H&N 1 was more heavily modulated with a corresponding increase in monitor units (MU). Chest 1 was a treatment of the chest wall following mastectomy of the right breast. This PTV had the largest volume (1451 cm$^3$) due to its substantial longitudinal length and additional involvement of the supraclavicular nodes, as well as extending more inferiorly than any other treatment. It was delivered with a 5.0 cm slice width. The final plan, Chest 2, treated the upper lobe of the right lung, and included portions of the mediastinum. All plans in this study were delivered with fixed jaw settings.

Table 1. Summary of clinical treatment plans delivered to the phantom

| Plan | Brain 1 | Brain 2 | H&N 1 (unilateral) | H&N 2 (bilateral) | Chest 1 | Chest 2 |
|---|---|---|---|---|---|---|
| PTV Volume (cm$^3$) | 364 | 12 | 445 | 811 | 1451 | 469 |
| PTV Prescribed Dose (Gy) | 60 | 54 | 54 | 70 | 50 | 60 |
| Fractions | 30 | 30 | 30 | 35 | 25 | 30 |
| PTV Dose per Fraction (Gy) | 2 | 1.8 | 1.8 | 2 | 2 | 2 |
| MU per Fraction | 3589 | 2928 | 5318 | 3968 | 5601 | 3946 |
| Slice Width (cm) | 2.5 | 1.0 | 2.5 | 2.5 | 5.0 | 2.5 |
| Pitch | 0.287 | 0.287 | 0.287 | 0.430 | 0.287 | 0.430 |
| Modulation Factor | 2.207 | 2.502 | 2.562 | 1.943 | 1.649 | 1.629 |
| Treatment Time (s) | 255 | 213 | 373 | 284 | 393 | 283 |
| Couch Travel (cm) | 10.8 | 3.4 | 18.3 | 14.8 | 29.2 | 15.3 |

PTV = planning target volume, MU = monitor units, H&N = head and neck.

**Phantom Irradiations**

An anthropomorphic Rando phantom (Radiology Support Devices, Long Beach, CA) consisting of 2.5 cm axial slices was selected to simulate a patient in the first trimester of pregnancy. The pelvis and abdomen of the phantom were replaced with 30 × 30 cm$^2$ virtual water blocks, as depicted in Figure 1. A PTW 30013 Farmer ionisation chamber connected to a PTW Unidos electrometer (PTW, Freiburg, Germany) was then inserted into the blocks such that the central electrode of the chamber was located at depth $d$ = 10 cm along the mid-line of the phantom. The chamber had a calibration traceable to a primary standards dosimetry laboratory, and each reading was corrected for temperature and pressure. Polarity and recombination effects were ignored.

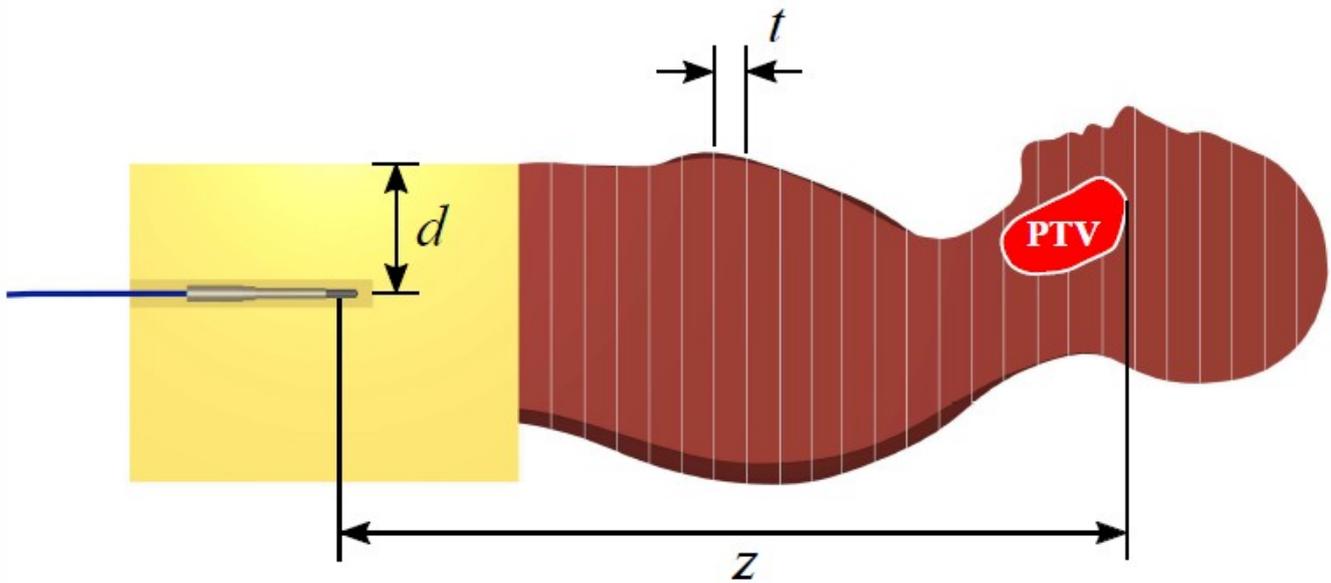

**Figure 1.** Sagittal cross-section of Rando phantom head and thorax with axial slice thickness $t$ = 2.5 cm. The abdomen has been replaced with virtual water blocks enclosing a Farmer ionisation chamber at depth $d$ = 10 cm. The distance between the measurement point and the superior boundary of the PTV, $z$, can be increased or decreased by adding or removing slices from the phantom.

The selected treatments were delivered to the phantom with an Accuray TomoTherapy Hi-Art treatment unit calibrated to the AAPM TG-51 protocol [11,12]. The separation between the measurement point and superior boundary of the PTV, $z$, corresponds to the maximum distance from the field edge. This distance was increased or decreased by adding or removing slices from the inferior thorax of the phantom. Each treatment was delivered at five values of $z$ to simulate a range of patient anatomy.

In contrast to conventional linear accelerators, the tomotherapy couch moves longitudinally through the bore at a constant rate during treatment. As such, the displacement of the measurement point relative to the edge of the field was steadily reduced during each delivery, corresponding to a total distance equal to the planned couch travel. In this study, the field edge is considered to be defined by the inferior jaw. Although the tomotherapy unit does have an MLC, it is binary in nature and so each leaf is either fully open or fully closed. In all cases, the measurement points never entered the primary field. Figure 2 illustrates the longitudinal displacement of the measurement point from the field edge for each treatment plan. Measurements were not made at points less than 5 cm from the field edge, as it should be possible to estimate the dose at such points using the treatment planning system in cases where the embryo or foetus is located so close to the PTV.

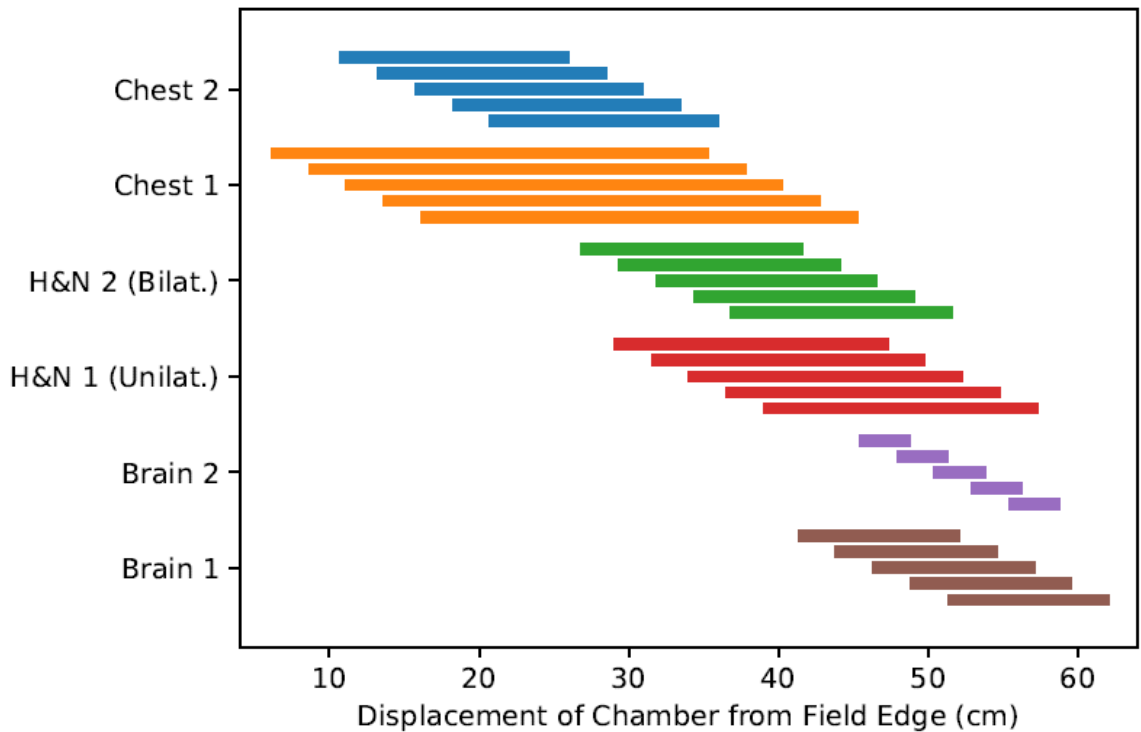

**Figure 2.** Position of the measurement point during each delivery. As the couch moves longitudinally during each treatment the ion chamber is drawn closer to the field.

## Results

Figure 3 presents the doses received for each treatment course as a function of the mean distance from the chamber to the inferior edge of the PTV (abbreviated to "mean chamber displacement"). The measured doses to the uterine region from all of the treatments were less than the 100 mGy established limit for significant risk in the first trimester [4].

However, the majority of treatments with mean chamber displacement less than 30 cm fell into the range of uncertain risk (50 – 100 mGy) [4]. All treatments with mean chamber displacement beyond 30 cm fell into the low risk category.

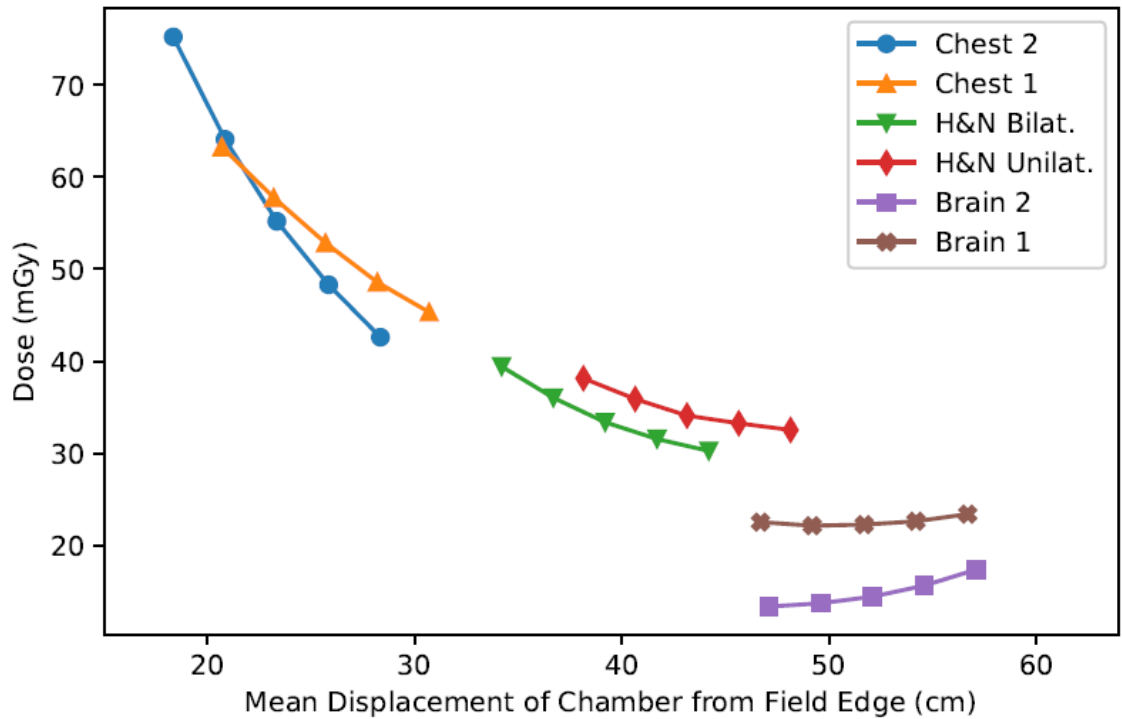

**Figure 3.** The dose delivered to the measurement point for all treatment plans.

The measurement results can be seen in Figure 4 expressed as a percentage of the prescribed dose to the PTV. The H&N 1 case had a larger relative dose than the H&N 2 case, potentially due to the increased MU and modulation. Chest 1 was also higher than Chest 2, and again had a higher MU, although this time the increased MU was necessary to treat the much longer volume and not due to increased modulation - indeed, the modulation factors for both plans were approximately equal. Brain 1 also showed a greater measured dose relative to the prescribed dose than Brain 2, with a corresponding increased MU. However, Brain 2 had a substantially smaller volume than Brain 1 and was also the only plan delivered with a 1 cm slice width. Measurements of both brain treatment plans tended to increase with increasing mean distance from the field edge.

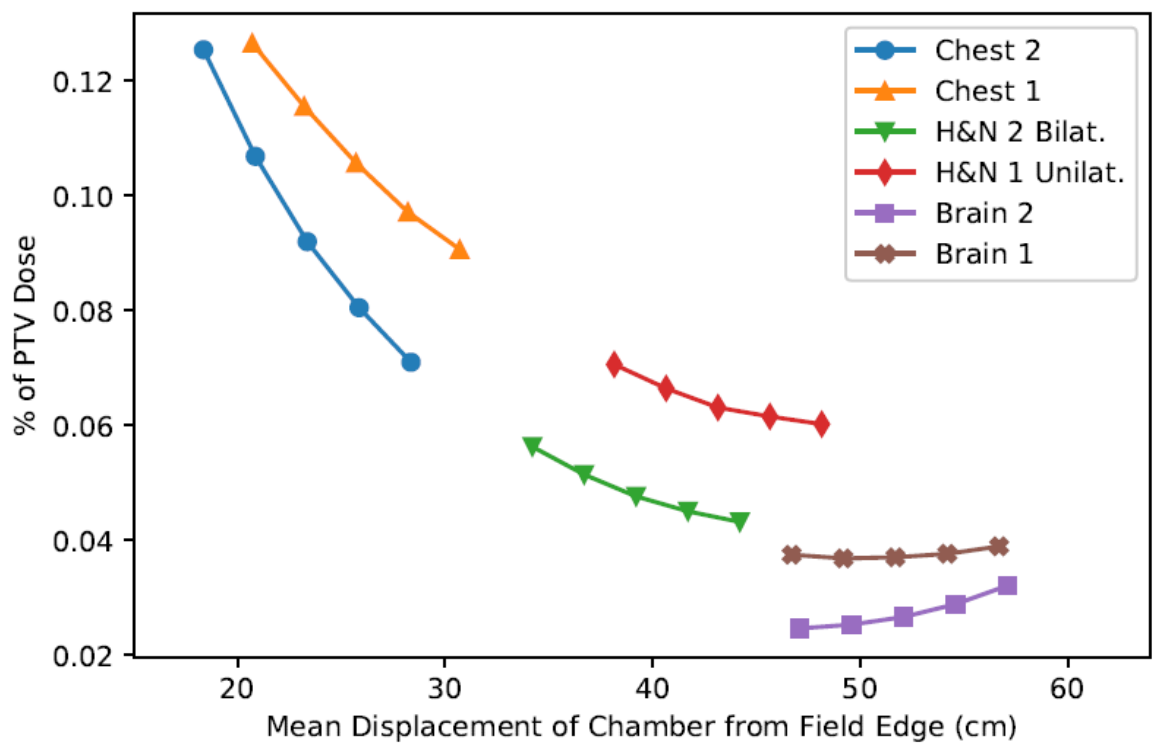

**Figure 4.** The dose delivered to the measurement point expressed as a percentage of the prescribed dose to the PTV.

To remove the potential influence of MU and total course dose, the data were renormalised in Figure 5 to express dose per MU for a single fraction, as most fractionation schemes have an approximate dose of 2 Gy per fraction. As can be seen in the figure, this resulted in all of the treatment plans that were delivered with a 2.5 cm slice thickness falling into a roughly exponential relationship. Only a single treatment plan was investigated at 1 cm and 5 cm slice thicknesses, so fits to these data were not attempted.

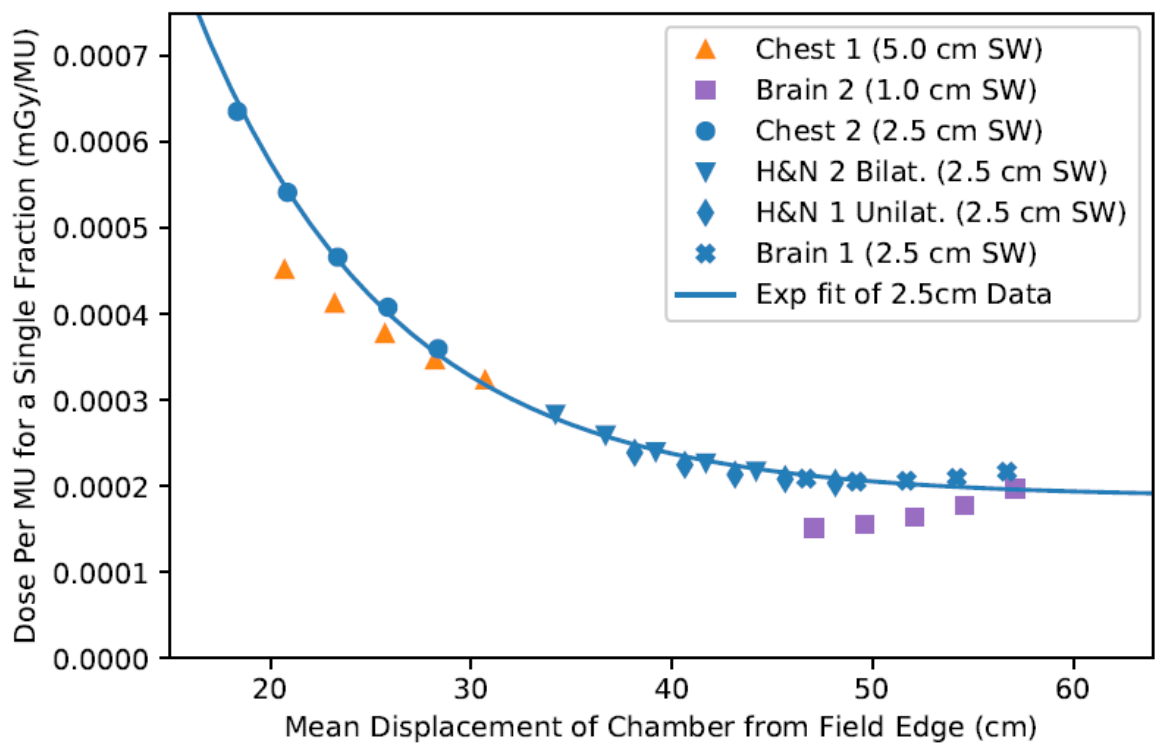

**Figure 5.** The dose per monitor unit measured in a single fraction. An exponential fit has been applied to the 2.5 cm slice width data.

## Discussion

No treatment plans considered in this study resulted in a dose to the embryo over the 100 mGy threshold for high risk biological effects [4]. The largest doses were encountered in the chest plans as they extended the most inferiorly, with the maximum measured dose being 75 mGy. These results compare favourably with earlier studies using 3D conformal radiotherapy (3DCRT). One study observed 38 mGy in a 48 Gy breast treatment [13], a second study recorded values in the range 21 – 76 mGy for breast treatments at various out-of-field distances [14], and a third study found first trimester doses of 50 and 83 mGy for treatments of the breast and lung respectively [15]. These results imply that tomotherapy may achieve similar foetal doses to 3DCRT for treatments of the chest. This is an interesting result given the increased MU and leakage dose typically associated with conventional intensity-modulated treatments (IMRT) [16,17,18].

An earlier study investigated a conventional five field IMRT head-and-neck treatment of a pregnant patient [19]. The patient was prescribed 68 Gy in 2 Gy fractions, with the shortest distance from the isocentre to the foetus being approximately 40 cm. Measurements returned results within the range 300 – 500 mGy, carrying a high enough risk that additional patient shielding was investigated. Comparatively, the tomotherapy measurements in the present study with similar geometries and prescriptions were a factor of 10 lower, in the range 30 – 40 mGy. This suggests that in the rare cases where modulated radiotherapy treatments of head-and-neck targets in pregnant patients are considered medically unavoidable, tomotherapy systems may provide a lower-dose alternative to conventional linac-based IMRT treatments.

The brain treatments investigated in this study resulted in the lowest foetal doses (13 – 25 mGy) due to being the furthest out-of-field. These values are smaller than those found in an earlier study which investigated a CyberKnife stereotactic IMRT treatment of a pregnant woman with a brain lesion, calculating an average foetal dose of 42 mGy [20]. Another study considered two 3DCRT brain cases and estimated doses of 30 mGy and 60 mGy, slightly larger than those found in this study [21]. This again implies that tomotherapy treatments lead to foetal doses that are on par with, or slightly lower than, those found in 3DCRT treatments.

However, an unexpected behaviour was observed in the measurements of the two brain treatment plans. In figures Figure 3 - Figure 5 it can be observed that the peripheral doses increased beyond a mean chamber displacement of approximately 50 cm. This might be explained by a thin spot in the head shielding at this angle, or perhaps due to the beam stopper preferentially scattering at this angle. This effect has not been seen in other tomotherapy studies as investigations of clinical plans have never measured out to this distance. Case-by-case measurements are advisable if dose estimates need to be performed at distances greater than 50 cm.

An exponential relationship between peripheral dose and mean chamber displacement was observed for the 2.5 cm slice width deliveries when the measurement was expressed as dose per MU for a single fraction. Knowledge of this relationship could be useful when dose estimates are required. The method for performing such an estimate would be:

1. Identify the point of interest (POI) for which the dose is to be estimated.
2. Calculate the distance from the POI to the superior aspect of the PTV.
3. Subtract half the planned couch travel from this value to reach the mean displacement of the POI from the field edge.
4. Using this mean displacement, interpolate from Figure 5 the dose per MU for a single fraction.
5. Multiply this value by both the MU per fraction and the number of fractions in the course, to arrive at the estimated dose for the full treatment course.

Note that this method holds only for 2.5 cm slice width treatment plans, and a dose per fraction of approximately 2 Gy is assumed. The result of this estimate may then inform the need for further out-of-field measurements or secondary estimation methods.

For some patients, a full course of radiation can take upwards of seven weeks, as was the case in the H&N 2 plan considered in this study. Depending on the trimester, foetal growth or movement would be expected throughout this time. To help account for this, peripheral doses for each plan were measured at five distances. It is evident from the results that neighbouring positions can have considerably different doses, particularly with measurement points closer to the treatment field. As such, it is recommended that the extent of this movement be considered on a case by case basis.

A clear limitation of this study was the selection of patient cases for inclusion. While these specific cases were selected as a representative sample of common sites treated in our local clinic, the results suggest that the peripheral dose is a complicated relationship involving PTV dimensions, field width, modulation, and other factors. It is therefore not clear how the peripheral dose would vary between two seemingly similar plans. Moreover, only a single anthropomorphic phantom was used, so the effect of substantially different patient anatomies has not been established. At the very least, the results of this study provide a general idea of the magnitude of peripheral doses to be expected.

## Conclusions

This study investigated foetal doses in brain, head-and-neck, and chest tomotherapy treatments. For treatments with typical prescribed doses (50 – 70 Gy) and where the mean foetal location is greater than approximately 30 cm from the field edge, avoidance of treatment is not justified. Additionally, this study provided a series of measurement data that may assist physicists in performing foetal dose estimations and inform clinicians when considering the use of tomotherapy to treat pregnant patients. This study has shown that tomotherapy has similar or decreased embryo or foetal dose compared to 3DCRT, but with an obvious increase in dosimetric quality commensurate with an IMRT treatment. Additionally, tomotherapy IMRT was shown to produce much lower embryo or foetal doses than five field IMRT with a conventional accelerator. Tomotherapy is therefore an attractive modality in cases where the treatment of a pregnant patient is unavoidable.

## Acknowledgements

This study was supported by a Royal Brisbane and Women's Hospital research grant.

## Conflicts of Interest

The authors report no conflicts of interest for this study.

x